\documentclass[12pt]{iopart}
\usepackage{iopams}
\usepackage{graphicx}
\usepackage{epsfig}
\usepackage{epstopdf}
\usepackage{cite}

\begin{document}

\title[]{Leaving bads provides better outcome than approaching goods in a social dilemma}

\author{Zhilong Xiao$^1$, Xiaojie Chen$^1$, and Attila Szolnoki$^2$}
\address{$^1$School of Mathematical Sciences, University of Electronic Science and Technology of China, Chengdu 611731, People's Republic of China}
\address{$^2$Institute of Technical Physics and Materials Science, Centre for Energy Research, P.O. Box 49, H-1525, Budapest, Hungary}
\ead{xiaojiechen@uestc.edu.cn}
\vspace{10pt}
\begin{indented}
\item[]
\end{indented}

\begin{abstract}
Individual migration has been regarded as an important factor for the evolution of cooperation in mobile populations. Motivations of migration, however, can be largely divergent: one is highly frustrated by the vicinity of an exploiter or defector, while other enthusiastically searches cooperator mates. Albeit both extreme attitudes are observed in human behavior, but their specific impacts on wellbeing remained unexplored. In this work, we propose an orientation-driven migration approach for mobile individuals in combination with the mentioned migration preferences and study their roles in the cooperation level in a two-dimensional public goods game. We find that cooperation can be greatly promoted when individuals are more inclined to escape away from their defective neighbors. On the contrary, cooperation cannot be effectively maintained when individuals are more motivated to approach their cooperative neighbors. In addition, compared with random migration, movement by leaving defectors can promote cooperation more effectively. By means of theoretical analysis and numerical calculations, we further find that when individuals only choose to escape away from their defective neighbors, the average distance between cooperators and defectors can be enlarged, hence the natural invasion of defection can be efficiently blocked. Our work, thus, provides further insight on how different migration preferences influence the evolution of cooperation in the unified framework of spatially social games.
\end{abstract}

\vspace{2pc}

\section{Introduction}
The emergence and maintenance of cooperation among unrelated individuals has been a puzzling phenomenon in nature and human societies~\cite{Axelrod_apsr81}.  Over the past decades, evolutionary game theory has provided a very competent framework for studying the evolution of cooperative behavior \cite{Axelrod_bb84,Nowak_n92,Nowak_s06,Santos_n08}. In particular, the public goods game has been recognized as a paradigm, which succinctly describes the essential dilemma of cooperation \cite{Nowak_hup06,Axelrod_bb84}. Recent works on the public goods game have proposed effective means to enable the evolution of cooperation, such as punishment \cite{Perc_njp12,Chen_sr15,Chen_pre15,Wang_amc18,He_amc19,Wang_sr13,Liu_mmmas19}, reward \cite{Szolnoki_njp12,Szolnoki_prsb15,Liu_csf18}, exclusion \cite{Szolnoki_pre17,Liu_sr17,Liu_c18,Quan_c19}, and individual migration~\cite{Perc_Bio10}.

Individual migration is an essential characteristic of living organisms \cite{Enquist_ab93}. It has been demonstrated that the mode of individual mobility does influence the evolutionary dynamics of cooperation among unrelated individuals, which has attracted intensive research activity
in recent years \cite{Pacheco_prl06,Vainstein_jtb07,Meloni_pre09,Sicardi_jtb09,Helbing_pnas09,Droz_Epjb09,Yang_pre10,Jiang_pre10,Aktipis_ehb11,Roca_pnas11,Zhang_pa11,Meloni_os17,Chen_y_pa16,Cheng_njp11,Cong_pone12,Fotouhi_rsif19,Takesue_epl19}. Theoretical and experimental studies have shown that individual mobility can promote the evolution of cooperation \cite{Cardillo_pre12,Chen_pre12,Vainstein_pre14,Wu_pre12,Fu_jsp13,Lewis_nc14,Wang_sr14,Antonioni_sr14,Takano_sr15,Wang_jtb15,Zhang_sr16,Cong_sr17,Avelino_pre18,Li_amc18,Ren_amc18,Chen_w_pa16,Li_pa19,Li_csf15,Cardinot_njp19,Vainstein_pa14}.
In particular, Meloni~et al. considered random migration for individuals playing the prisoner's dilemma game on a two-dimensional plane and found that cooperation can be maintained when the moving velocity of individuals is not too high \cite{Meloni_pre09}. Subsequently, Cardillo~et al. \cite{Cardillo_pre12} found that in the public goods game, played on a two-dimensional plane, low mobility promotes cooperation, whereas high velocity can disrupt cooperation. What is more, Helbing and Yu \cite{Helbing_pnas09} proposed success-driven migration under which individuals move to the location which is surrounded by cooperators in the prisoner's dilemma game on a square lattice and demonstrated that such mode of migration leads to the outbreak of cooperation. On the other hand, Chen et al. proposed risk-driven migration in the collective-risk social dilemma game on a square lattice and found that risk-driven migration dramatically enhances the evolution of public cooperation when individuals move away from unfavorable locations \cite{Chen_pre12}.

It is worth pointing out that most of previous works consider random migration \cite{Meloni_pre09,Cardillo_pre12}, success-driven migration \cite{Helbing_pnas09}, or risk-driven migration \cite{Chen_pre12} separately. They do not consider the orientation-driven migration under which different migration preferences or migration modes are considered in a unified framework. Indeed individuals can adjust their moving orientation according to these preferred modes. By means of migration they can move away from unfavorable environment, pursuit the profitable circumstances, or choose other directions as they wish in realistic situations. However, it is still unclear how such orientation-driven migration influences the evolution of cooperation and which mode of individual migration can promote the evolution of cooperation more effectively.

In this work, we thereby propose an orientation-driven migration into a population of mobile individuals playing the public goods game. We assume that individuals can choose the direction of mobility depending on the strategy types of their neighbors on a two-dimensional plane under such orientation-driven migration. Correspondingly, individuals can choose to escape from neighboring defectors or move to neighboring cooperators according to the settings of orientation parameters. By means of Monte Carlo simulations and numerical calculations, we show that cooperation can be best maintained when individuals choose to escape from neighboring defectors, when the mobility velocity is not too high. On the contrary, cooperation cannot be promoted when individuals are more inclined to move to neighboring cooperators. Furthermore, compared with random migration, we find that escaping from neighboring defectors can better promote the evolution of cooperation.

\section{Model}
In our model, we consider a population of $N$ individuals who play the public goods game on a two-dimensional plane of linear size $L$ with periodic boundary conditions. Hence, the density of individuals is defined as $\rho= N/L^{2}$. Each individual is described via position and velocity vectors on the two-dimensional plane. Initially each individual is distributed at random in the plane via using two independent random variables from $[0, L]$ interval, and correspondingly individual $i$'s initial position is assigned as $\mathbf{r}_{i}(0)=[{x}_{i}(0), {y}_{i}(0)]$.

Once the initial configuration of the system is set, two dynamical processes coevolve: orientation-driven migration and strategy evolution. By adopting Ref.~\cite{Angelani_prl12}, at every time step we assume that each individual $i$ moves with a constant speed $v$ and its position $\mathbf{r}_{i}(t)=[{x}_{i}(t), {y}_{i}(t)]$ and velocity are updated by means of the following equations
\begin{eqnarray}
\mathbf{r}_{i}(t+1) = \mathbf{r}_{i}(t)+\mathbf{v}_{i}(t+1), \label{eq1}
\end{eqnarray}
\begin{eqnarray}
\mathbf{v}_{i}(t+1) = v\mathbf{\widehat{v}}_{i}(t), \label{eq2}
\end{eqnarray}
where we used $\Delta t=1$ and
$\mathbf{\widehat{v}}_{i}(t)$ is a unit vector which is determined by the following equation
\begin{eqnarray}
\mathbf{v}_{i}(t) = \eta \mathbf{\widehat{f}}_{i}^{(CD)}(t) + \mu\mathbf{g}_{i}(t). \label{eq3}
\end{eqnarray}
\indent The first term in the right side of Eq.~($3$) describes the orientation-driven force for individual $i$ by strategy distribution among the neighbors, and $\eta$ quantifies its relative strength. For simplicity without loss of generality, $\eta$ is set to one in this study. Furthermore, we assume that
\begin{eqnarray}
\mathbf{f}_{i}^{(CD)} = (1-\beta)\mathbf{\widehat{f}}_{i}^{(C)}+\beta\mathbf{\widehat{f}}_{i}^{(D)}, \label{eq4}
\end{eqnarray}
where
\begin{eqnarray}
\mathbf{f}_{i}^{(C)} = -\displaystyle\sum_{j\in S_i^{(C)}}h(r_{ij})\mathbf{\widehat{r}}_{ij} \label{eq5}
\end{eqnarray}
and
\begin{eqnarray}
\mathbf{f}_{i}^{(D)} = \displaystyle\sum_{j\in S_i^{(D)}}h(r_{ij})\mathbf{\widehat{r}}_{ij}. \label{eq6}
\end{eqnarray}
Here the sum of Eq.~($5$) [Eq.~($6$)] is over individual $i$'s neighboring cooperators (defectors) $j$ who are within an Euclidean distance less than the threshold distance of interaction $R$ that is, $\sqrt{[x_{i}(t)-x_{j}(t)]^{2}+[y_{i}(t)-y_{j}(t)]^{2}}\le R$. Here $h(r)$ is a weight function and is set as $r^{-w}$, where $w>1$ in agreement with Ref.~\cite{Angelani_prl12}. Notably,
$0\le\beta\le1$ is a key parameter of our model characterizing the relative weight between the two extreme motivation attitudes. For $\beta=0$, individual $i$ concentrates to go closer to neighboring cooperators. Whereas for $\beta=1$, individual $i$ focuses exclusively
to escape away from neighboring defectors.

\indent The second term in the right side of Eq.~($3$) describes a steric repulsive force, so that individual overlap can be prevented. The related $\mu$ parameter quantifies its
relative strength on $\mathbf{v}_{i}(t)$. We consider that
\begin{equation}
\mathbf{g}_{i} = \displaystyle\sum_{j\in S_i^{(rep.)}}\mathbf{g}(\mathbf{r}_{i}-\mathbf{r}_{j}), \label{eq7}
\end{equation}
where the sum is over neighbors within a sphere of radius $R$ surrounding individual $i$. In agreement with Ref.~\cite{Angelani_prl12} the function $\mathbf{g}$ is set as
\begin{equation}
\mathbf{g}(\mathbf{r}) = \frac{\mathbf{\widehat{r}}}{1+\exp[(r-r_{f})/\sigma]}, \label{eq8}
\end{equation}
where $r =|\mathbf{\widehat{r}}|$, $r_{f}$ describes the length scale of repulsion, and $\sigma$ describes the steepness.

The second ingredient of our dynamical model is the evolutionary public goods game played by mobile individuals. Initially an individual $i$ is designated as a cooperator $[s_{i}(0)=1]$ or a defector $[s_{i}(0)=0]$ with equal probability. At each time step, we consider that the neighborhood of a given individual $i$ is made up by all the individuals $j$ who are within an Euclidean distance less than the threshold distance of interaction $R$~\cite{Boccaletti_pr06}. In other words, when $\sqrt{[x_{i}(t)-x_{j}(t)]^{2}+[y_{i}(t)-y_{j}(t)]^{2}}\le R$, individuals $i$ and $j$ are connected at time step $t$ and we have $A_{ij}(t)=1$, otherwise $A_{ij}(t)=0$ in the adjacency matrix $A(t)$. Evidently, we have $A_{ii} = 0$. Importantly, individual $i$ whose number of neighbors is $n_{i}$ does not only play a single public goods game with all its corresponding neighbors, but also plays the public goods games in alternative groups where its neighbors are the focal players. In a public goods game where individual $i$ participates in, each cooperator contributes the same cost $c$ ($c$ is set to $1$ in this study without loss of generality), while each defector contributes nothing. The total contribution from cooperators is multiplied by a multiplication factor $r$ and then distributed equally among all group members independently of their strategies, hence the total payoff of individual $i$ at time step $t$ is given by
\begin{eqnarray}
P_{i}(t)=\displaystyle\sum_{j=1}^{N}[A_{ij}(t)+\delta_{ij}]\frac{\displaystyle\sum_{k=1}^{N}[A_{jk}(t)+\delta_{jk}]s_{k}(t)c r}{n_{j}(t)+1}-[n_{i}(t)+1]s_{i}(t)c, \label{eq9}
\end{eqnarray}
where $\delta_{ij}=1$ if $i=j$, otherwise $\delta_{ij}=0$.

After each round, each individual $i$ has a chance to imitate the strategy of a randomly chosen neighbor $j$. If $P_{j}(t) < P_{i}(t)$, no update occurs. Otherwise, the strategy transfer occurs with the probability
\begin{eqnarray}
q = \frac{P_{j}(t)-P_{i}(t)}{M}, \label{eq10}
\end{eqnarray}
where $M$ ensures the proper normalization and is given by the maximum possible difference between the total payoffs of individuals $i$ and $j$ \cite{Santos_n08}. We note that this strategy update rule is also known as discrete replicator rule~\cite{Iranzo_plosone12}.

We emphasize that during the evolutionary process, there exist complicated coupling effects between the evolutionary dynamics of individuals' motions and strategies. In particular, individuals' motion can change the interaction structures of the mobile population, which can also influence the strategy evolution in the population. On the other hand, individuals' strategy updates can also have consequences on how neighboring players move. We correspondingly study this coevolutionary model by means of Monte Carlo simulations. Simulations are carried out in a population with size $N=1000$. As the key quantity, the fraction of cooperators is defined as the density of cooperators in the whole population. We are interested in concentrating on how the mobility speed $v$, the orientation-driven weight $\beta$, the strength of steric repulsive force $u$, and the threshold distance $R$ influence the fraction of cooperators, in order to clearly explore the effects of our proposed orientation-driven migration on the evolution of cooperation. To do that, we set $r=5.75$, $\rho=2$, $r_f=0.2$, $\sigma=0.1$, and $w=2$ for simplicity. We find that our main results remain valid when these parameter values are changed. In addition, when the above described updating rules are applied, the mobile population may converge to one of the two possible absorbing states, which are full cooperation or full defection. To gain representative behavior we run $200$ independent realizations for each set of parameter values and compute the fraction of times that the population evolves to full cooperation. Alternatively, if the population does not converge to an absorbing state after $10^6$ updates, then the cooperation level is determined in the stationary state by averaging the fraction of cooperators in the population over the last $10^4$ updates.

\section{Results}

\begin{figure}
\centering
\includegraphics[width=3.5 in]{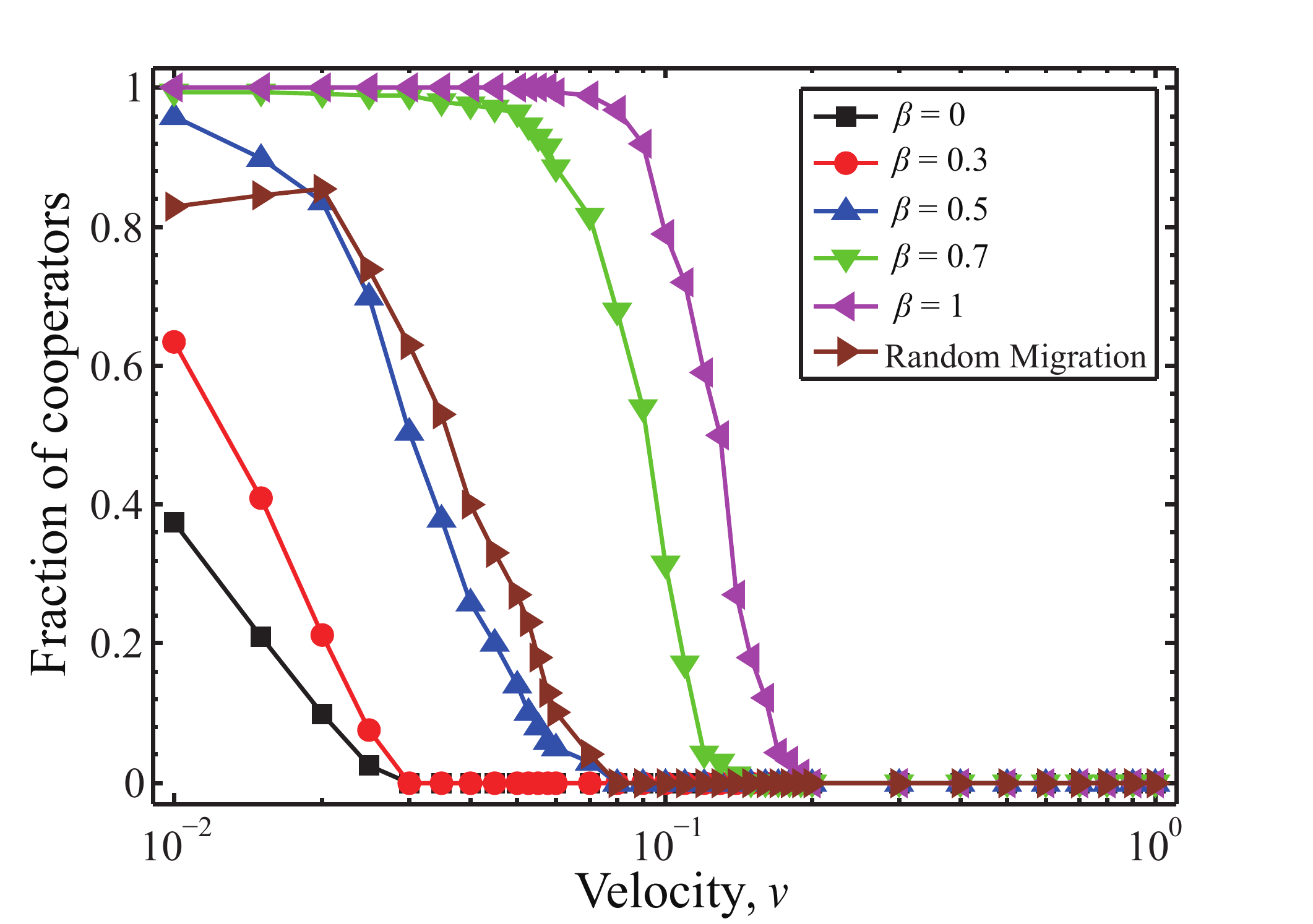}
\caption{Fraction of cooperators in dependence on the speed $v$ for  random migration and orientation-driven migration with different values of $\beta$. Parameters: $\mu=1$ and $R=1$.}
\label{fig1}
\end{figure}

We first present the fraction of cooperators in dependence on the mobility speed $v$ for different values of the orientation-driven weight $\beta$, as shown in Fig.~\ref{fig1}. We find that for each value of $\beta$ the fraction of cooperators decreases with increasing the speed $v$, but cooperators can flourish for low values of $v$. In particular, the highest level of cooperation can be reached for large values of $\beta$. For the sake of comparison, we also show the fraction of cooperators as a function of the mobility speed $v$ for random migration in Fig.~\ref{fig1}. We find that when the speed is not high, the fraction of cooperators for orientation-driven migration with large $\beta$ is always higher than that for random migration. In addition, note that the cooperation level for $\beta=0.5$ is close to that obtained for random migration when the same value of speed $v$ is applied. This may be because the case of $\beta=0.5$ corresponds to a situation where attraction by cooperators has the same strength to the aversion to defectors, which approaches the case of random migration in which diffusion is independent of strategies of neighboring players~\cite{Vainstein_pre14}.

\begin{figure}
\centering
\includegraphics[width=3.5 in]{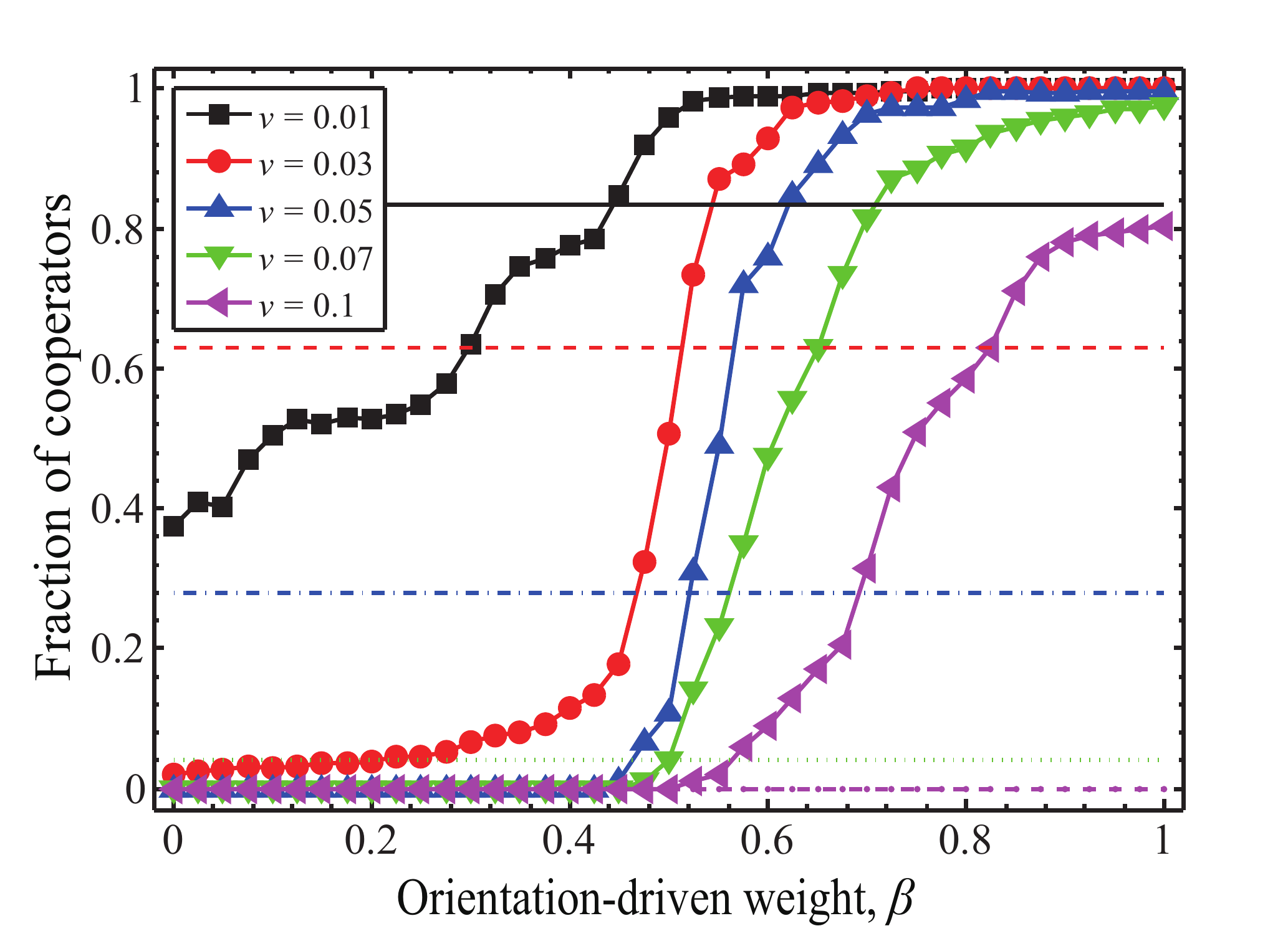}
\caption{Fraction of cooperators as a function of the orientation-driven weight $\beta$ for different values of speed $v$. Dashed line are used to indicate the fractions of cooperators for random migration at these speed values, i.e.,
black ($v=0.01$), red ($v=0.03$), blue ($v=0.05$), green ($v=0.07$), and purple ($v=0.1$). Parameters: $R=1$ and $\mu=1$.}
\label{fig2}
\end{figure}

In order to qualify the effect of orientation-driven weight $\beta$ on the evolution of cooperation in detail, we further show the fraction of cooperators as a function of $\beta$ for different values of $v$, as shown in Fig.~\ref{fig2}. We find that for each value of $v$ the fraction of cooperators increases gradually as the value of $\beta$ increases. Notably, full cooperation can be reached for large $\beta$, especially when the mobility speed is low. For small $\beta$ values, however, the cooperation promoting effect is moderate even at low mobility speed. In addition, Fig.~\ref{fig2} shows the comparison of random migration case with the orientation-driven cases obtained at different speed values.
We can see that for each value of $v$, there exists a critical value of $\beta$, above which orientation-driven migration can better promote cooperation than random migration. These results
indicate that when individuals are more inclined to escape away from defectors in their interactive neighborhoods, the evolution of cooperation can be promoted. In particular, cooperation can be best promoted when individuals concentrate exclusively to escape away from their neighboring defectors. On the contrast, the evolution of cooperation is not supported
when individuals are focusing to move close to cooperators in their neighbors.

\begin{figure*}
\centering
\includegraphics[width=6.4in]{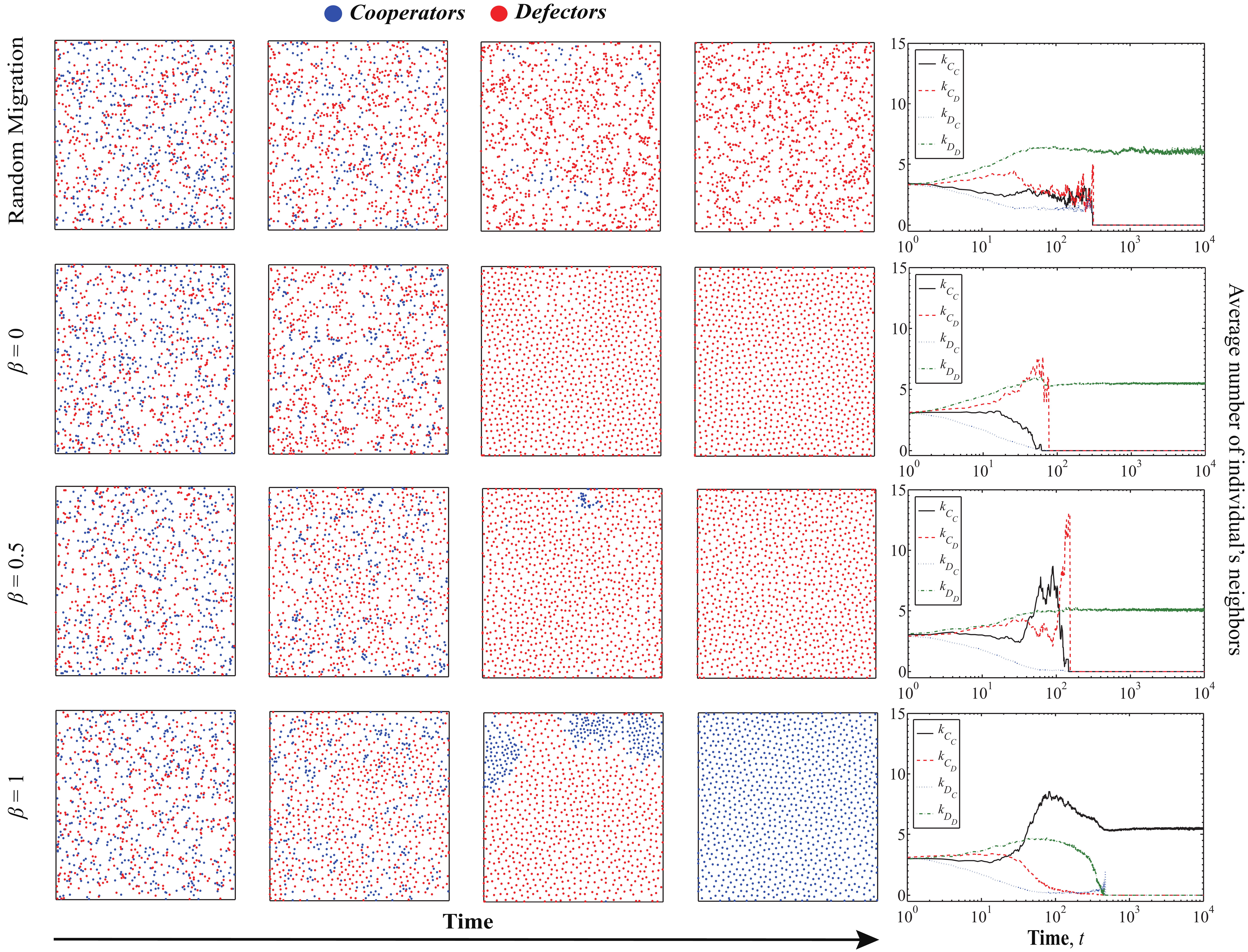}
\caption{First four columns depict the time evolution of spatial patterns for random migration and orientation-driven migration with three different values of $\beta$. Here blue color represents cooperator while red color marks defector players. The fifth column depicts how the average number of individual's neighbors change in time in the mentioned cases.
Here, $k_{C_{C}}$ ($k_{C_{D}}$) denotes the average number of neighboring cooperators (defectors) around cooperator players,
while $k_{D_{C}}$ ($k_{D_{D}}$) marks the average number of neighboring cooperators (defectors) of defector players. We define here $k_{D_{C}} = 0$ and $k_{D_{D}} = 0$ when there are no defectors in the population, whereas $k_{C_{C}} = 0$ and $k_{C_{D}} = 0$ mean that there are no cooperators in the population. Parameters are $v=0.1$, $R=1$, and $\mu=1$.}
\label{fig3}
\end{figure*}

In order to gain deeper insight about the effects of orientation-driven movement on the evolution of cooperation, we present a series of snapshots of strategy evolution about the microscopic process for three representative values of $\beta$ in Fig.~\ref{fig3}. For the sake of comparison, we also plot the typical snapshots of strategy evolution for random migration here. Meanwhile, we further illustrate
how the average number of a cooperator's or defector's neighboring cooperators or defectors changes in time for these different cases, as presented in the rightmost column of Fig.~\ref{fig3}. We can see that for random migration (top row), widespread cooperative patches occur in the mobile population at the early stage of evolution. With the invasion of defectors, then several isolated cooperators and tiny separated cooperator formations are found in the two-dimensional plane. We can also find that with the decrease of the number of cooperators in the population, the average numbers of neighboring cooperators and defectors of cooperators both decrease during this period of evolution. Finally, cooperators will disappear and instead defectors will dominate the whole population. When the orientation-driven migration is considered, we can find that for $\beta=0$ (second row), although cooperators move towards to neighboring cooperators, meanwhile defectors also move towards to neighboring cooperators. Due to the evolutionary advantage of defectors to cooperators and such kind of orientation-driven migration for moving close to cooperators, the average number of neighboring cooperators of cooperators decreases, while the average number of neighboring defectors of cooperators increases during the period of evolution. Correspondingly, cooperators' clusters cannot be formed, and cooperators will disappear soon in the population. While for $\beta=0.5$ (third row), on one hand individuals will consider to move close to their neighboring cooperators, on the other hand they will consider to escape away from their neighboring defectors. Correspondingly, during the evolutionary process a cooperators' cluster can be gradually formed from widespread cooperative patches. However, the cluster size is not large enough, so it cannot resist the invasion of defectors successfully. During this period of evolution, we can see that the average number of neighboring cooperators of cooperators can first increase, but with the invasion of defectors it will decrease then. Meanwhile the average number of neighboring defectors of cooperators increases. Subsequently, the cooperators' cluster will shrink, and finally disappear. For $\beta=1$ (bottom row), individuals concentrate to escape away from their neighboring defectors. We can find that a single large compact cluster of cooperators can be formed from numerous cooperators patches in the two-dimensional plane. Correspondingly, the average number of neighboring cooperators of cooperator players can increase, while the average number of neighboring defectors of cooperator agents decreases during the period of evolution. Consequently, this compact cluster can not just resist the invasion of defectors, but it can also grow and expand. As a result, cooperation will finally prevail in the whole population.

The comparison of time evolution in the fifth column highlights that $k_{D_{D}}$ always grows first due to the successful imitation of defector strategy in the mixed initial state. But this effect is weakened significantly at a large $\beta$ value where players (including cooperators) are motivated to escape from the vicinity of defectors. Consequently, this is the only case where $k_{C_{D}}$ decays in time, hence defectors are not fed anymore by neighboring cooperators. This explains the striking difference between the outcomes of plotted cases.

\begin{figure}
\centering
\includegraphics[width=6in]{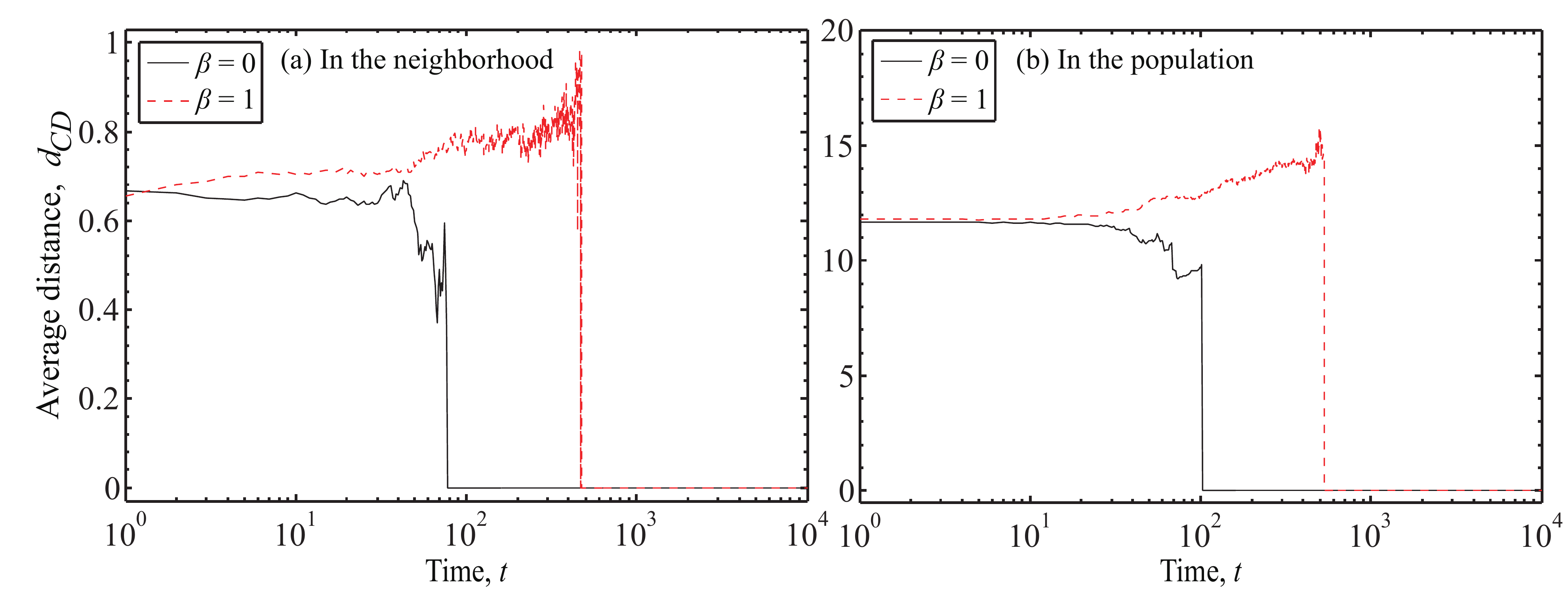}
\caption{Panel (a) shows the average distance between cooperators and defectors as a function of time calculated among neighboring players for $\beta=0$ and $\beta=1$. Panel (b) shows the time evolution of the average distance between cooperators and defectors calculated in the whole population for the same $\beta$ values. Parameters are $v=0.1$, $R=1$, and $\mu=1$.}
\label{fig4}
\end{figure}

To support our argument quantitatively, in Fig.~\ref{fig4}~(a) we show how the average distance $d_{CD}$ between cooperators and defectors in the neighborhoods evolves in time for $\beta=0$ and $\beta=1$. We can observe that in the early stage the average distance between neighboring cooperators and defectors gradually increases with time for $\beta=1$. In contrast, for $\beta=0$ the same average distance remains practically unchanged in the beginning and decays later. We note, however, that this late decay is just a simple consequence of the fact that the population becomes homogeneous where only defectors remain. In addition,
in the intermediate state when both strategies are present
the average distance between neighboring cooperators and defectors is always higher for $\beta=1$ than that for $\beta=0$.

For comparison we also show how the average distance $d_{CD}$ between cooperators and defectors in the population evolves in time for $\beta=0$ and $\beta=1$, as presented in Fig.~\ref{fig4}(b). We can find that in the whole population the average distance of cooperator and defector players remains unchanged at the early stage of evolution independently of the value of $\beta$. But later this average distance gradually increases for $\beta=1$, while it decreases for $\beta=0$. Furthermore, the mentioned critical distance for $\beta=1$ always exceeds the same value for $\beta=0$. These results demonstrate that the motivation to escape away from neighboring defectors can widen effectively the average distance between cooperators and defectors: both in the neighborhoods and in the whole population. This effect, however, is completely missing, when players are motivated to approach cooperator neighbors. Hence, we can conclude that
the evolution of cooperation can be better promoted by escaping away from defectors than searching the vicinity of cooperators.

In the following, we present a simple model calculation to explain further the paramount importance of above described average distance of competing strategies. Accordingly, we consider two simplified mathematical models, which respectively describe the motion among one cooperator and two defectors (Appendix~A) and the motion among two cooperators and one defector
player (Appendix~B).

\begin{figure}[h!]
\centering
\includegraphics[width=6in]{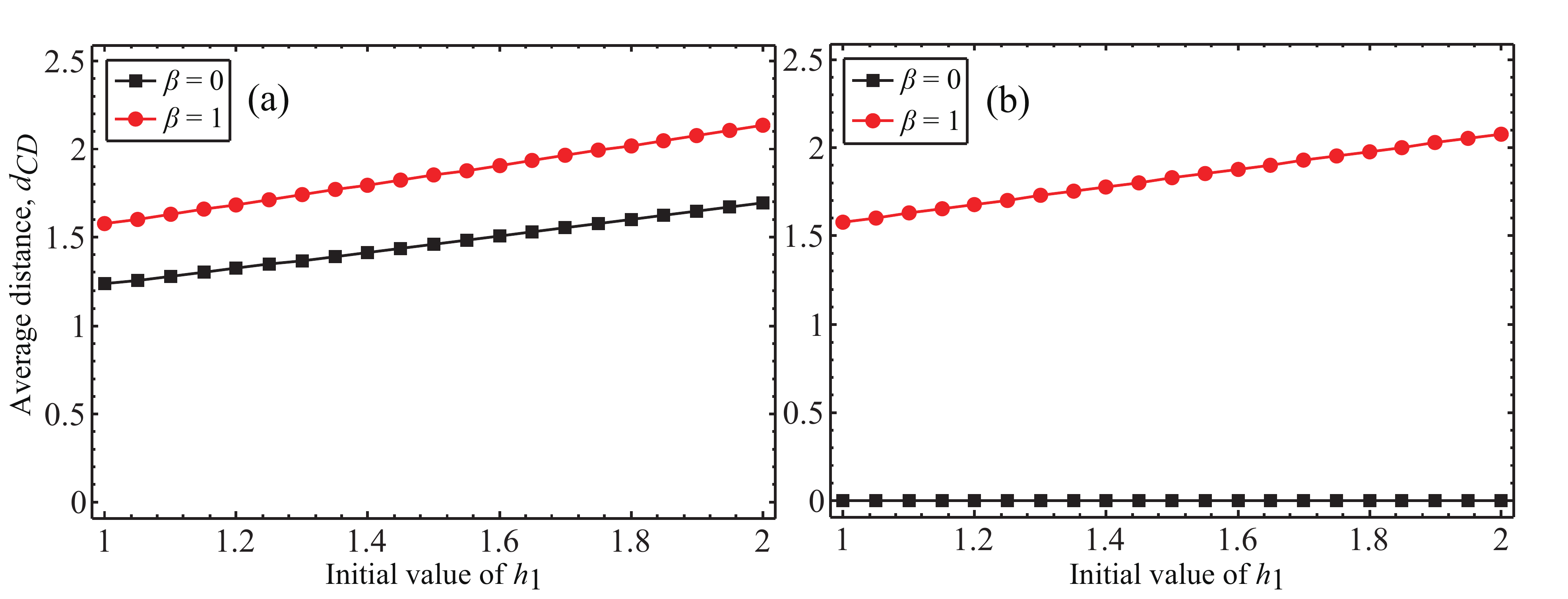}
\caption{Panel (a) shows the average distance between cooperators and defectors in the simplified motion model of one cooperator and two defectors as a function of the initial value of $h_1$ for $\beta=0$ and $\beta=1$. Panel (b) shows the average distance between cooperators and defectors in the simplified motion model of one defector and two cooperators as a function of the initial value of $h_1$ for $\beta=0$ and $\beta=1$. Parameters: $v=0.1$, $R=1$, and $\mu=1$.}
\label{fig5}
\end{figure}

These models allow us to obtain dynamical equations of motion for $\beta=0$ and $\beta=1$ extreme cases (see Appendix~A and B for more details). Here we define the average distance between cooperators and defectors as $d_{CD} = (h_{1}+h_{2})/2$, where $h_1$ is the distance between cooperator $C$ ($C_1$) and defector $D_1$ ($D$) in Appendix~A (B), and $h_2$ is the distance between cooperator $C$ ($C_2$) and defector $D_2$ ($D$) in Appendix~A (B). By means of numerical calculations, we present the average distance $d_{CD}$ as a function of the initial value $h_1(0)$ for $\beta=0$ and $\beta=1$ in these simplified models, as depicted in Fig.~\ref{fig5}. In Fig.~\ref{fig5}~(a), we can find that the average distance $d_{CD}$ increases with the initial value of $h_1$ both for $\beta=0$ and $\beta=1$. But for each initial value of $h_1$, the average distance $d_{CD}$ for $\beta=1$ is always higher than that for $\beta=0$. In Fig.~\ref{fig5}(b), we can find that the average distance $d_{CD}$ increases with the initial value of $h_1$ for $\beta=1$, and the average distance for $\beta=0$ is always zero for each initial value of $h_1$. Hence the former is always higher than the latter for each initial value of $h_1$. Indeed these motion patterns of our simplified models can also appear in our model, and hence they clearly explain why the distance between cooperators and defectors is widened when players are principally motivated to leave defector neighbors and may be reduced when players are focusing to approach cooperator neighbors. This difference, as we stressed, has a decisive factor on the final evolutionary outcome.

\begin{figure}
\centering
\includegraphics[width=3.5 in]{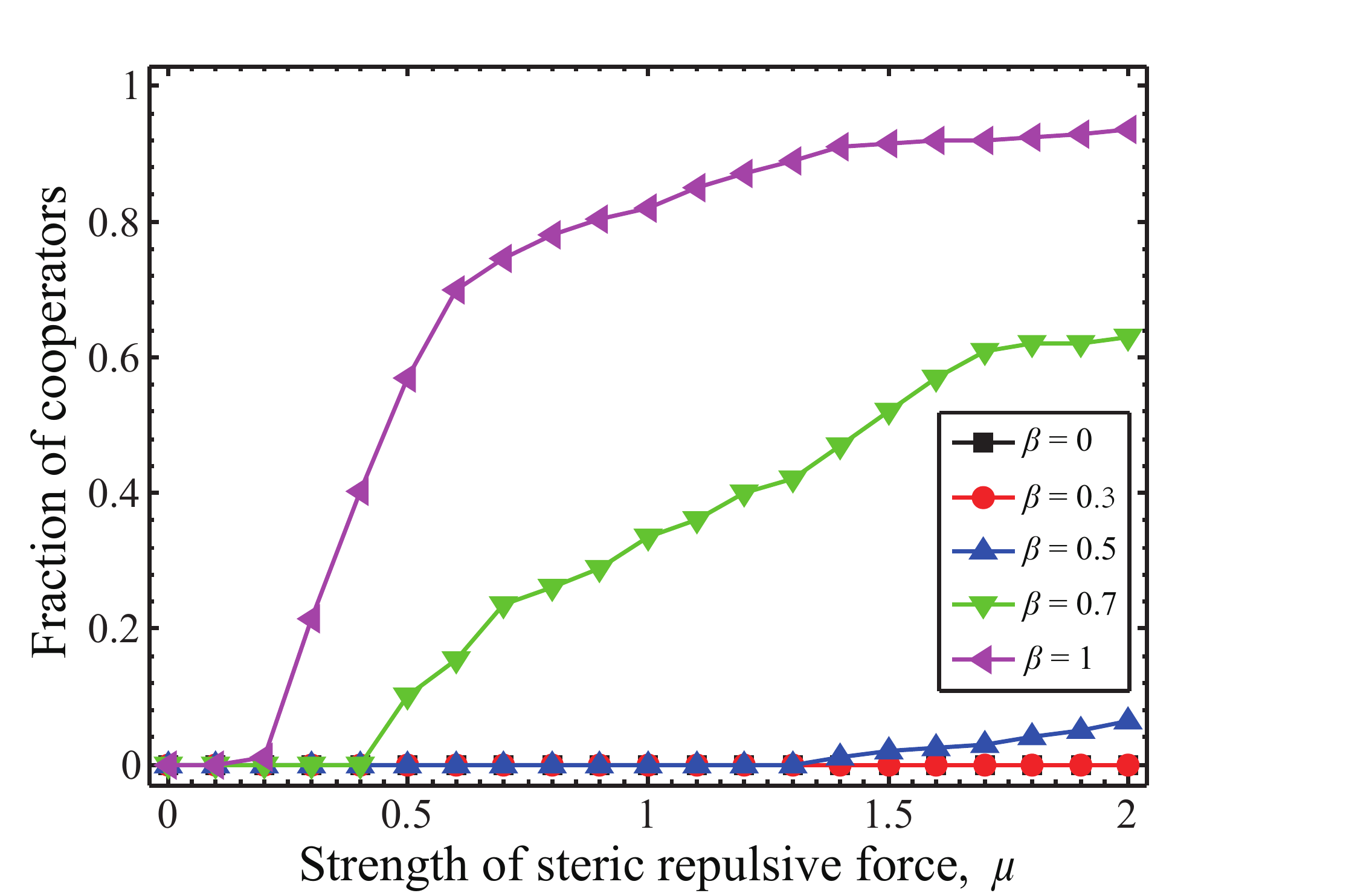}
\caption{Fraction of cooperators in dependence on the strength of steric repulsive force $\mu$ for different values of $\beta$ as indicated in the legend. Parameters are $v=0.1$ and $R=1$. We note that random migration yields zero cooperation level at these parameter values.}
\label{fig6}
\end{figure}

In what follows, we study the influence of the strength $\mu$ of steric repulsive force on the evolution of cooperation for different values of $\beta$. Our results are summarized in Fig.~\ref{fig6} where we plot the fraction of cooperators in dependence on $\mu$. We see that the cooperation level can always be raised by increasing the value of $\mu$ especially at high $\beta$ values. In the absence of relevant repulsive force at small $\mu$ values, however, cooperators cannot survive. We note that in the random migration case the applied parameter values would also result in a full defector state. These findings indicate that the introduction of the steric repulsive force can promote the evolution of cooperation under orientation-driven migration with high value $\beta$.

Finally, it remains of interest to explore how the threshold distance of interactions $R$ influences the evolution of cooperation for different values of $\beta$ under the orientation-driven migration protocol. As shown in Fig.~\ref{fig7}, we observe that the cooperation level always decreases if we increase the interaction range. This effect is specially pronounced at small $\beta$ values. Furthermore, when higher mobility speed is applied (not shown) the decay of cooperation level is even more stressful. Notably, the cooperation level for random migration is less than that for large $\beta$ value. These findings support that orientation-driven migration outperforms random migration for the evolution of cooperation at large values of $\beta$ and small values of $R$.

\begin{figure}
\centering
\includegraphics[width=3.5 in]{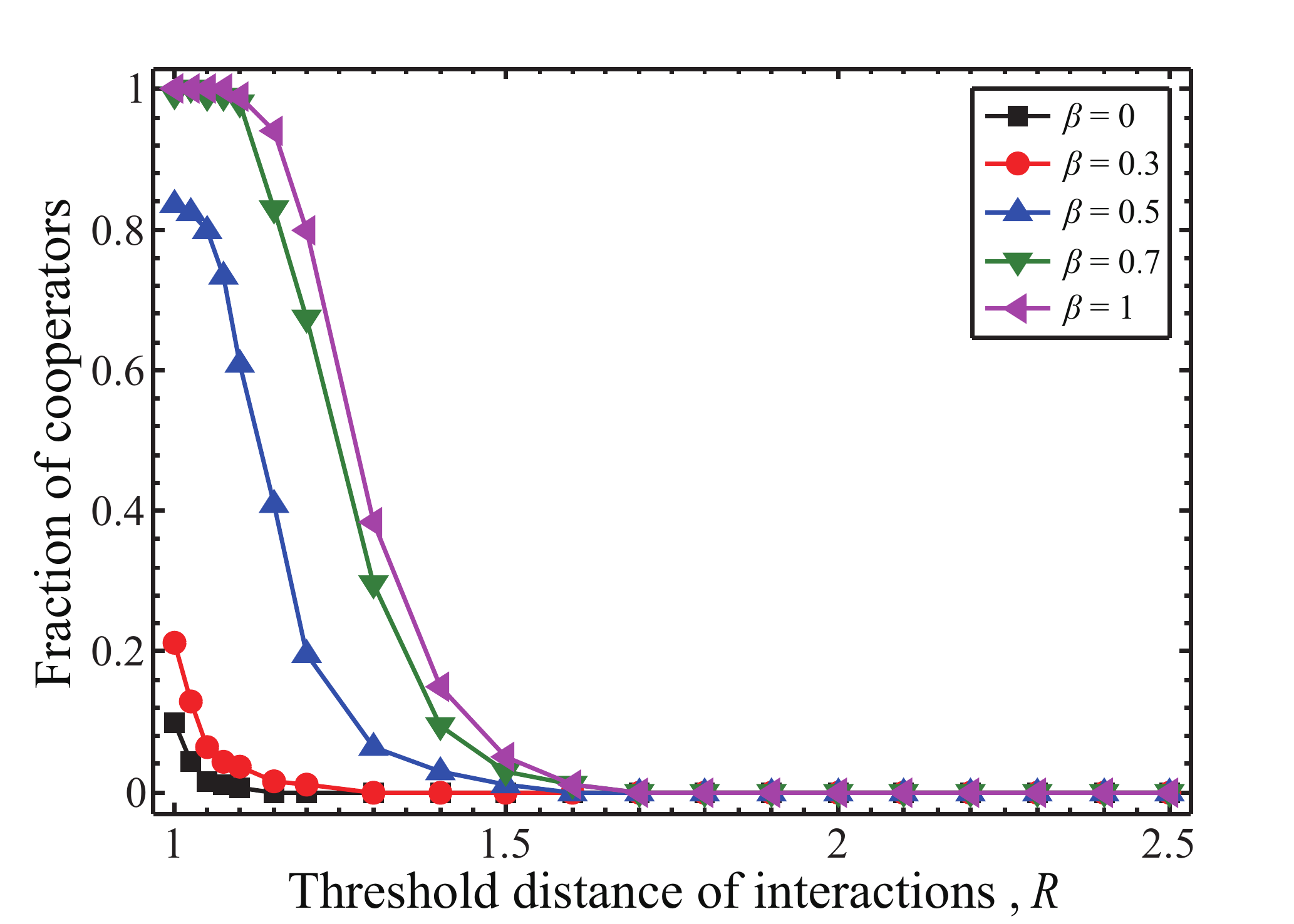}
\caption{Fraction of cooperators in dependence on the threshold distance of interaction range $R$ for different values of $\beta$ as indicated in the legend. Parameters are $v=0.02$ and $\mu=1$. We note that random migration would result in less cooperation level at these parameter values, when compared with the case of large $\beta$ value.}
\label{fig7}
\end{figure}

\section{Discussion}
In this work, we have proposed an orientation-driven migration approach into the spatial public goods game and studied how it influences the evolution of cooperation. Under the orientation-driven migration, each individual can adjust its motion direction according to the motion directions of its neighbors. In principle, individuals prefer to move closer to their neighboring cooperators or favor to escape away from their neighboring defectors. Considering these two extreme
driving forces into the orientation-driven migration approach,
in the framework of Monte Carlo simulations
we have found that the orientation-driven migration can strongly enhance the evolution of cooperation when the speed of individuals is not too high. In particular, cooperation can be promoted when individuals are more inclined to evade defectors in their neighbors, whereas cooperation cannot be effectively maintained when individuals are more inclined to move close to cooperators in their neighbors. Furthermore,
compared with random migration, escaping away from neighboring defectors for individuals can promote the evolution of cooperation more effectively. By means of theoretical analysis and numerical calculations, we further find that the key feature is the average distance of competing strategies, hence
escaping away from neighboring defectors can widen the average distance between cooperator and defector individuals. This quantity has a paramount importance, because its large value
can effectively block the invasion of defectors into cooperators and hence is favorable to the formation and expansion of cooperative clusters for the evolution of cooperation. In addition, we have found that cooperation can be more enhanced by high strength of steric repulsive force and low threshold distance of interaction.

The importance of our observation is based on the fact that
individual migration is pervasive in living organisms, and has been considered into evolutionary game models \cite{Helbing_pnas09,Cardillo_pre12,Chen_pre12,Vainstein_pre14}. It has been found that it can lead to the outbreak of cooperation. In particular, when individual migration preferences are considered, moving away unfavorable environment \cite{Chen_pre12} and moving into profitable circumstance  \cite{Helbing_pnas09} can be regarded as two significantly different migration modes for individuals. Previous works have demonstrated that these options may both greatly promote the evolution of cooperation \cite{Helbing_pnas09,Chen_pre12}. But if these two different driving forces for individual migration are both considered into the same framework of spatial games, which mode of individual migration can promote the evolution of cooperation more? Our work has clearly answered this question, and we have found that cooperation can be best maintained when individuals only choose to escape away from their neighboring defectors. On the contrary, cooperation cannot be effectively maintained when individuals only choose to move close to their neighboring cooperators. Furthermore, we find that our proposed orientation-driven migration approach can promote cooperation for low mobility, which is similar to the finding in Ref.~\cite{Cardillo_pre12} that
observed low mobility promotes cooperation under random migration. However, compared with random migration studied in Ref.~\cite{Cardillo_pre12}, we find that escaping from neighboring defectors can better promote the evolution of cooperation. Our work may thus unveil the evolution of cooperation driven by different migration preferences, and we hope that this research will contribute relevantly to our understanding of their role in determining the ultimate fate of the mobile population.

\section*{Acknowledgments}
This research was supported by the National Natural Science Foundation of China (Grant Nos. 61976048 and 61503062), by the Fundamental Research Funds of the Central Universities of China, and by the Hungarian National Research Fund (Grant K-120785).

\appendix
\section{Simplified motion model of one cooperator and two defectors}
In this paper, we consider a simplified motion model in which there are one cooperator $C$ and two defectors $D_{1}$ and $D_{2}$ , and aim to derive the dynamical equations in the scenario where the weight function $h(r)$ is a power-law function. To do that, we first set the position and velocity of cooperator $C$ as $\mathbf{r}_{C}=(x_{C}, y_{C})$ and $\mathbf{v}$, respectively. And we set the position and velocity of defector $D_{1}$ as $\mathbf{r}_{D_1}=(x_{D_{1}}, y_{D_{1}})$ and $\mathbf{v}_{1}$, respectively; the position and velocity of defector $D_{2}$ as $\mathbf{r}_{D_2}=(x_{D_{2}}, y_{D_{2}})$ and $\mathbf{v}_{2}$, respectively. Correspondingly, we have $| \mathbf{v} |$ = $| \mathbf{v}_{1} |$ = $| \mathbf{v}_{2} | =v$. We further have $\mathbf{v}_{i}= v\hat{\mathbf{V}}_{i}$ and $\mathbf{v}= v\hat{\mathbf{V}}$, where $\hat{\mathbf{V}}_{i}$ and $\hat{\mathbf{V}}$ are the unit vectors, and $i=1, 2$. Furthermore, we denote with $\mathbf{h}_{1}=\mathbf{r}_{C}-\mathbf{r}_{D_1}$ ($\mathbf{h}_{2}=\mathbf{r}_{C}-\mathbf{r}_{D_2}$) be the vector distance between cooperator $(C)$ and defector $D_{1}$ ($D_{2}$).  Correspondingly, we have $\mathbf{h}_{i}=h_i\hat{\mathbf{h}}_{i}$, where $\hat{\mathbf{h}}_{i}$ is the unit vector and $h_{i}$ is the distance between cooperator $C$ and defector $D_{i}$. In addition, we have $h_{i}=|\mathbf{h}_{i}|=\sqrt{(x_{C}-x_{D_{i}})^{2}+(y_C-y_{D_{i}})^2}$, where $i=1, 2$.

\section*{A1.The case of $\beta=0$}

\begin{figure}
\setcounter{figure}{0}
\renewcommand\thefigure{{A1}}
\centering
\includegraphics[width=3.0 in]{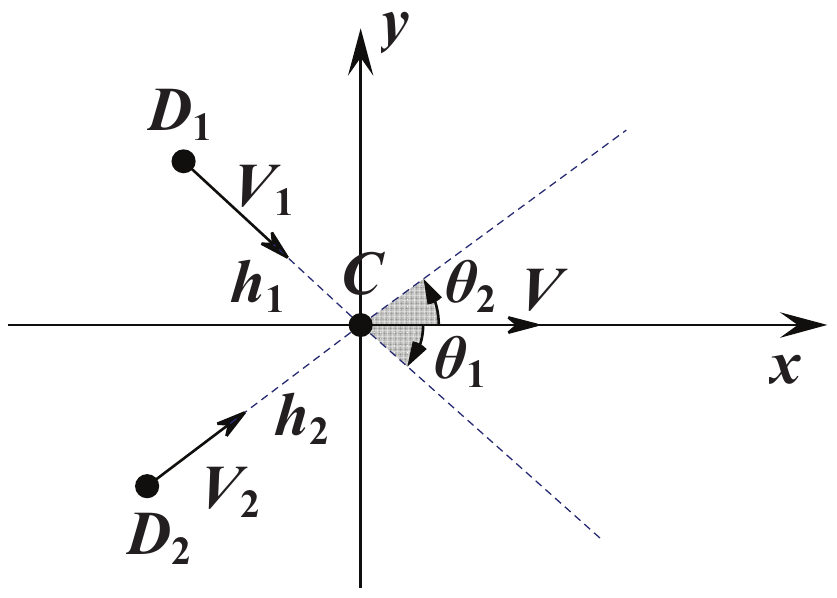}
\caption{Two defectors $D_1$ and $D_2$ move close to cooperator $C$. $\theta_{1} $ ($\theta_{2} $) represents the angle between the vector velocities of cooperator $(C_{1})$ and defector $D_{1}$ ($D_{2})$.}
\label{FigA1}
\end{figure}

In this case, we assume that two defectors move towards cooperator $C$ directly. Since there are no other cooperators in the neighborhood of cooperator $C$, we consider that cooperator $C$ moves along with a fixed direction, which is the direction of $\mathbf{v}$. For simplicity, but without losing generality, we consider that the migration direction of cooperator $C$ is the positive direction of $x$-axis in the cartesian coordinate (see Fig.~\ref{FigA1}), and correspondingly set the dynamical position of cooperator $C$ as $(vt, 0)$. Hence, we have $h_{i}=\sqrt{(vt-x_{D_{i}})^{2}+y^2_{D_{i}}}$.

Meanwhile, for defectors $D_{1}$ and $D_{2}$ the dynamical equations of motion can be described as
\begin{eqnarray}
\left\{
\begin{array}{lcl}
\dot{x}_{D_{i}} = v\cos\theta_{i},\\
\dot{y}_{D_{i}} = v\sin\theta_{i},
\end{array}
\right.
\end{eqnarray}
where $\theta_{i}$ $(i=1,2)$ represents the angle between the vector velocities of cooperator $C$ and defector $D_{i}$. For simplicity but without losing generality, we consider that
$x_{D_1}(0)=-h_1(0)\cos\theta_1(0)$ and $y_{D_1}(0)=-h_1(0)\sin\theta_1(0)$ for defector $D_1$, and $x_{D_2}(0)=-h_2(0)\cos\theta_2$ and $y_{D_2}(0)=-h_2(0)\sin\theta_2(0)$ for defector $D_2$.

In addition, we have $\tan\theta_{i}=\frac{-y_{D_{i}}}{vt-x_{D_{i}}}$. Accordingly, we have
\begin{eqnarray}
\tan^{2}\theta_{i}=\frac{y^2_{D_{i}}}{(vt-x_{D_{i}})^{2}}, \nonumber
\end{eqnarray}
and
\begin{eqnarray}
\frac{1}{\cos^{2}\theta_{i}}=1+\frac{y^2_{D_{i}}}{(vt-x_{D_{i}})^{2}}.  \nonumber
\end{eqnarray}

Considering that $\theta_{i}$ should be restricted between $(0, \pi/2)$ or $(-\pi/2, 0)$, thus we have
\begin{eqnarray}
\cos\theta_{i}=\frac{vt-x_{D_{i}}}{\sqrt{(vt-x_{D_{i}})^{2}+y^2_{D_{i}}}},  \nonumber
\end{eqnarray}
and
\begin{eqnarray}
\sin\theta_{i}=\frac{-y_{D_{i}}}{\sqrt{(vt-x_{D_{i}})^{2}+y^2_{D_{i}}}}.  \nonumber
\end{eqnarray}

Hence, the dynamical equations of motion for defectors become
\begin{eqnarray}
\left\{
\begin{array}{lcl}
\dot{x}_{D_{i}} = \frac{v^2t-vx_{D_{i}}}{\sqrt{(vt-x_{D_{i}})^{2}+y^2_{D_{i}}}},\\
\dot{y}_{D_{i}} = \frac{-vy_{D_{i}}}{\sqrt{(vt-x_{D_{i}})^{2}+y^2_{D_{i}}}}.
\end{array}
\right.
\end{eqnarray}

According to the above equations, we can further calculate the average distance $d_{CD}$ between cooperator $C$ and the two defectors $D_1$, $D_2$ for $\beta=0$ in this simplified motion model. To do that, we solve Eq. (A.2) via numerical integrations
by using Runge-Kutta method \cite{Press_cup92} with time step $dt=10^{-3}$. The initial conditions are $h_{2}(0) = 2$, $-\pi/2< \theta_1(0)<0$, $0 < \theta_2(0)<\pi/2$, and $1\leq h_1(0)\leq 2$. Then we can respectively obtain the $h_1$ and $h_2$ values, and correspondingly have $d_{CD} = (h_{1}+h_{2})/2$. We emphasize that for each initial value $h_1(0)$, we can obtain a $d_{CD}$ value for fixed $\theta_1(0)$ and $\theta_2(0)$ values, and the average distance $d_{CD}$ for $\beta=0$ in Fig.~\ref{fig5}(a) is obtained by averaging over all these distance values for uniformly distributed initial values $\theta_1(0)$ between $(-\pi/2, 0)$ and uniformly distributed initial values $\theta_2(0)$ between $(0, \pi/2)$.

\section*{A2. The case of $\beta=1$}
\begin{figure}
\setcounter{figure}{0}
\renewcommand\thefigure{{A2}}
\centering
\includegraphics[width=3.0 in]{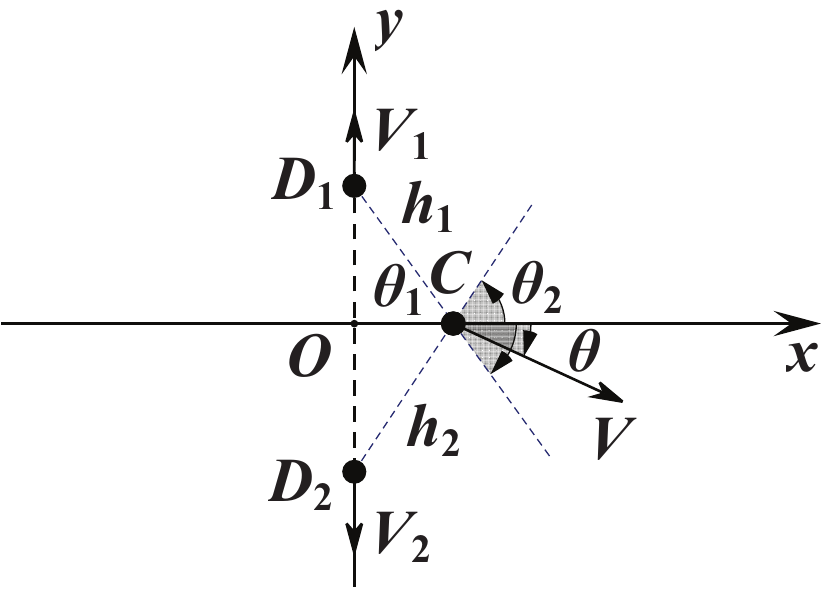}
\caption{Individuals escape away from defectors in the situation in which there are one cooperator and two defectors. $\theta$ represents the angle between the positive $x$-axis and the vector velocity of cooperator $C$, and $\theta_{i}$ represents the angle between the positive $x$-axis and the vector $\mathbf{h}_{i}$.}
\label{FigA2}
\end{figure}

In this case, we assume that defectors $D_1$ and $D_2$ will move with the opposite directions. For simplicity but without losing generality, we assume that the migration direction of defector $D_{1}$ is the positive direction of $y$-axis, while the migration direction of defector $D_{2}$ is the negative direction of $y$-axis (see Fig.~\ref{FigA2}). We further set the positions of two defectors as $(0, y_{1}+vt)$ and $(0, y_{2}-vt)$, respectively, where $y_1$ and $y_2$ respectively represent the initial values of $y_{D_{1}}$ and $y_{D_2}$. Hence we have
$y_1=-h_1(0)\sin\theta_1(0)$ and $y_2=-h_2(0)\sin\theta_2(0)$. For simplicity, we assume that initially cooperator $C$ locates on the positive $x$-axis. Correspondingly, we have $x_C(0)=h_1(0)\cos\theta_1(0)$ and $y_C(0)=0$.

For cooperator $C$, the dynamical equations of motion can be described as
\begin{eqnarray}
\left\{
\begin{array}{lcl}
\dot{x}_{C} = v \cos\theta,\\
\dot{y}_{C} = v \sin\theta,
\end{array}
\right.
\end{eqnarray}
where $\theta$ represents the angle between the vector velocity of cooperator $C$ and the positive $x$-axis.

In addition, we have $\tan\theta_{i}=\frac{y_{C}-y_{D_{i}}}{x_{C}}$. We further have
\begin{eqnarray}
\frac{1}{\cos^{2}\theta_{i}}=1+\frac{{(y_{C}-y_{D_{i}})^2}}{x_{C}^{2}}. \nonumber
\end{eqnarray}

We consider that $\theta_{i}$ should be restricted between $(0, \pi/2)$ or $(-\pi/2, 0)$ and the position of the cooperator $x_{C}\geq0$, so we have
\begin{eqnarray}
\cos\theta_{i}=\frac{x_{C}}{\sqrt{x_{C}^{2}+(y_{C}-y_{D_{i}})^2}},  \nonumber
\end{eqnarray}
and
\begin{eqnarray}
\sin\theta_{i}=\frac{y_{C}-y_{D_{i}}}{\sqrt{x_{C}^{2}+(y_{C}-y_{D_{i}})^2}}.  \nonumber
\end{eqnarray}

In addition, the direction of cooperator $C$ is given by
\begin{eqnarray}
\hat{\mathbf{V}}=\frac{h_{1}^{-w}\hat{\mathbf{h}}_{1}+h_{2}^{-w}\hat{\mathbf{h}}_{2}}{| h_{1}^{-w}\hat{\mathbf{h}}_{1}+h_{2}^{-w}\hat{\mathbf{h}}_{2}|}, \nonumber
\end{eqnarray}
where $\mathbf{V} = h_{1}^{-w}\hat{\mathbf{h}}_{1}+h_{2}^{-w}\hat{\mathbf{h}}_{2}$. We further have
\begin{eqnarray}
| \mathbf{V} | = \sqrt{h_{1}^{-2w}+h_{2}^{-2w}+2h_{1}^{-w}h_{2}^{-w}\cos(\theta_{2}-\theta_{1})}. \nonumber
\end{eqnarray}

Using the definition of the inner product of vectors, we obtain
\begin{eqnarray}
\left\{
\begin{array}{lcl}
\cos(\theta-\theta_{1}) = \hat{\mathbf{V}}\cdot\hat{\mathbf{h}}_{1}=\frac{h_{1}^{-w}+h_{2}^{-w}\cos(\theta_{2}-\theta_{1})}{\sqrt{h_{1}^{-2w}+h_{2}^{-2w}+2h_{1}^{-w}h_{2}^{-w}\cos(\theta_{2}-\theta_{1})}},\\
\cos(\theta_{2}-\theta) =\hat{\mathbf{V}}\cdot\hat{\mathbf{h}}_{2} = \frac{h_{2}^{-w}+h_{1}^{-w}\cos(\theta_{2}-\theta_{1})}{\sqrt{h_{1}^{-2w}+h_{2}^{-2w}+2h_{1}^{-w}h_{2}^{-w}\cos(\theta_{2}-\theta_{1})}}. \nonumber
\end{array}
\right.
\end{eqnarray}

As a result, we have
\begin{eqnarray}
\left\{
\begin{array}{lcl}
\cos\theta = \frac{h_{1}^{-w}\sin\theta_{2}+h_{2}^{-w}\cos(\theta_{2}-\theta_{1})\sin\theta_{2}-h_{2}^{-w}\sin\theta_{1}-h_{1}^{-w}\cos(\theta_{2}-\theta_{1})\sin\theta_{1}}{\sqrt{h_{1}^{-2w}+h_{2}^{-2w}+2h_{1}^{-w}h_{2}^{-w}\cos(\theta_{2}-\theta_{1})}\sin(\theta_{2}-\theta_{1})},\\
\sin\theta = \frac{h_{1}^{-w}\cos\theta_{2}+h_{2}^{-w}\cos(\theta_{2}-\theta_{1})\cos\theta_{2}-h_{2}^{-w}\cos\theta_{1}-h_{1}^{-w}\cos(\theta_{2}-\theta_{1})\cos\theta_{1}}{\sqrt{h_{1}^{-2w}+h_{2}^{-2w}+2h_{1}^{-w}h_{2}^{-w}\cos(\theta_{2}-\theta_{1})}\sin(\theta_{1}-\theta_{2})}. \nonumber
\end{array}
\right.
\end{eqnarray}

Hence, the dynamical equations of the motion for cooperator $C$ are given as
\begin{eqnarray}
\left\{
\begin{array}{lcl}
\dot{x}_{C} = v\frac{h_{1}^{-w}\sin\theta_{2}+h_{2}^{-w}\cos(\theta_{2}-\theta_{1})\sin\theta_{2}-h_{2}^{-w}\sin\theta_{1}-h_{1}^{-w}\cos(\theta_{2}-\theta_{1})\sin\theta_{1}}{\sqrt{h_{1}^{-2w}+h_{2}^{-2w}+2h_{1}^{-w}h_{2}^{-w}\cos(\theta_{2}-\theta_{1})}\sin(\theta_{2}-\theta_{1})},\\
\dot{y}_{C} = v\frac{h_{1}^{-w}\cos\theta_{2}+h_{2}^{-w}\cos(\theta_{2}-\theta_{1})\cos\theta_{2}-h_{2}^{-w}\cos\theta_{1}-h_{1}^{-w}\cos(\theta_{2}-\theta_{1})\cos\theta_{1}}{\sqrt{h_{1}^{-2w}+h_{2}^{-2w}+2h_{1}^{-w}h_{2}^{-w}\cos(\theta_{2}-\theta_{1})}\sin(\theta_{1}-\theta_{2})},
\end{array}
\right.
\end{eqnarray}
where $h_{1}=\sqrt{x_{C}^{2}+(y_{C}-y_{1}-vt)^2}$, $h_{2}=\sqrt{x_{C}^{2}+(y_{C}-y_{2}+vt)^2}$, $\cos\theta_{i}=\frac{x_{C}}{\sqrt{x_{C}^{2}+(y_{C}-y_{D_{i}})^2}}$, and $\sin\theta_{i}=\frac{y_{C}-y_{D_{i}}}{\sqrt{x_{C}^{2}+(y_{C}-y_{D_{i}})^2}}$.

According to the above equations, we can further calculate the average distance $d_{CD}$ between cooperator $C$ and two defectors $D_1$, $D_2$ for $\beta=1$ in this simplified motion model. To do that, we solve Eq.~(A.4) via numerical integrations
by using Runge-Kutta method \cite{Press_cup92} with time step $dt=10^{-3}$. Here the initial conditions are $h_{2}(0) = 2$, $w=2$, $-\pi/2 < \theta_1(0)<0$, $0 < \theta_2(0)<\pi/2$, and $1\leq h_1(0)\leq 2$. Then we can respectively obtain the $h_1$ and $h_2$ values and calculate $d_{CD} = (h_{1}+h_{2})/2$. We emphasize that for each initial value $h_1(0)$, we can obtain a $d_{CD}$ value for fixed $\theta_1(0)$ value, and the average distance $d_{CD}$ for $\beta=1$, plotted in Fig.~\ref{fig5}(a),
is obtained by averaging over all these distance values for uniformly distributed initial values $\theta_1(0)$ between $(-\pi/2, 0)$ and uniformly distributed initial values $\theta_2(0)$ between $(0, \pi/2)$.

\section{Simplified motion model of one defector and two cooperators}
In the following, we consider a simplified motion model in which there are a defector $(D)$ and two cooperators $C_{1}$ and $C_{2}$, and then derive the dynamical equations in the scenario where the weight function $h(r)$ is a power-law function. To do that, we first set the position and velocity of defector $D$ as $\mathbf{r}_{D}=(x_{D}, y_{D})$ and $\mathbf{v}$, respectively. And we set the position and velocity of cooperator $C_{1}$ as $\mathbf{r}_{C_1}=(x_{C_{1}}, y_{C_{1}})$ and $\mathbf{v}_{1}$, respectively; the position and velocity of cooperator $C_2$ as $\mathbf{r}_{C_2}=(x_{C_{2}}, y_{C_{2}})$ and $\mathbf{v}_{2}$, respectively. Correspondingly, we have
$|\mathbf{v}|$ = $|\mathbf{v}_{1} |$ = $| \mathbf{v}_{2}| = v$. We further have $\mathbf{v}_{i}= -v\hat{\mathbf{V}}_{i}$ and $\mathbf{v}= -v\hat{\mathbf{V}}$, where $\hat{\mathbf{V}}_{i}$ and $\hat{\mathbf{V}}$ are the unit vectors, and $i=1, 2$. Furthermore, we denote by
$\mathbf{h}_{1}=\mathbf{r}_{D}-\mathbf{r}_{C_1}$ ($\mathbf{h}_{2}=\mathbf{r}_{D}-\mathbf{r}_{C_2}$) the distance vector of cooperator $C_{1}$ ($C_{2}$) and defector $D$. Correspondingly, we have $\mathbf{h}_{i}=h_i\hat{\mathbf{h}}_{i}$, where $\hat{\mathbf{h}}_{i}$ is the unit vector and $h_{i}$ is the distance between defector $D$ and cooperator $C_{i}$. In addition, we have $h_{i}=\sqrt{(x_{D}-x_{C_{i}})^{2}+(y_D-y_{C_{i}})^2}$, where $i=1, 2$.

\section*{B1.The case of $\beta=0$}
\begin{figure}
\setcounter{figure}{0}
\renewcommand\thefigure{{B1}}
\centering
\includegraphics[width=3.0 in]{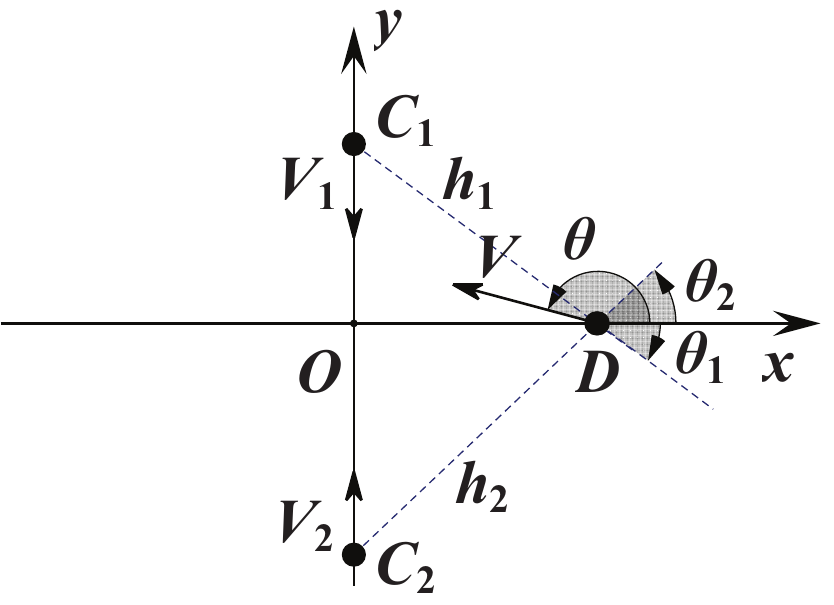}
\caption{Individuals move close to cooperators in the situation in which there are two cooperators and one defector. $\theta$ represents the angle between the positive $x$-axis and the vector velocity of defector $D$, and $\theta_{i}$ represents the angle between the positive $x$-axis and the vector $\mathbf{h}_{i}$.}
\label{FigB1}
\end{figure}

In this case, we assume that cooperators $C_1$ and $C_2$ will move towards to each other. For simplicity without losing generality, we assume that the migration direction of cooperator $C_1$ is the negative direction of $y$-axis, while the migration direction of cooperator $C_{2}$ is the positive direction of $y$-axis (see Fig.~\ref{FigB1}). We further set the positions of two cooperators as $(0, y_{1}-vt)$ and $(0, y_{2}+vt)$, respectively. Hence we have $y_1=-h_1(0)\sin\theta_1(0)$ and $y_2=-h_2(0)\sin\theta_2(0)$. For simplicity but without losing generality, we assume that initially defector $D$ locates on the positive $x$-axis. Correspondingly, we have $x_D(0)=h_1(0)\cos\theta_1(0)$ and $y_D(0)=0$.

For defector $D$, the dynamical equations of motion can be thus
described as
\begin{eqnarray}
\left\{
\begin{array}{lcl}
\dot{x}_{D} = v\cos\theta,\\
\dot{y}_{D} = v\sin\theta,
\end{array}
\right.
\end{eqnarray}
where $\theta$ represents the angle between the vector velocity of cooperator $C$ and the positive $x$-axis.

In addition, we have $\tan\theta_{i}=\frac{y_{D}-y_{C_{i}}}{x_{D}}$. We further have
\begin{eqnarray}
\frac{1}{\cos^{2}\theta_{i}}=1+\frac{{(y_{D}-y_{C_{i}})^2}}{x_{D}^{2}}.  \nonumber
\end{eqnarray}

We consider that when defector $(D)$ moves close to the origin $(O)$, the direction of defector $(D)$ becomes the negative or positive direction of $y$-axis, so we have $x_{D}\geq0$. Considering that $\theta_{i}$ should be restricted between $(0, \pi/2)$ or $(-\pi/2, 0)$, thus we have
\begin{eqnarray}
\cos\theta_{i}=\frac{x_{D}}{\sqrt{x_{D}^{2}+(y_{D}-y_{C_{i}})^2}},  \nonumber
\end{eqnarray}
and
\begin{eqnarray}
\sin\theta_{i}=\frac{y_{D}-y_{C_{i}}}{\sqrt{x_{D}^{2}+(y_{D}-y_{C_{i}})^2}}.  \nonumber
\end{eqnarray}

In addition, the direction of defector $D$ is given by
\begin{eqnarray}
\hat{\mathbf{v}}=-\frac{h_{1}^{-w}\hat{\mathbf{h}}_{1}+h_{2}^{-w}\hat{\mathbf{h}}_{2}}{|h_{1}^{-w}\hat{\mathbf{h}}_{1}+h_{2}^{-w}\hat{\mathbf{h}}_{2}|}. \nonumber
\end{eqnarray}
where $\mathbf{V} = -h_{1}^{-w}\hat{\mathbf{h}}_{1}-h_{2}^{-w}\hat{\mathbf{h}}_{2}$. We further have
\begin{eqnarray}
|\mathbf{V} | = \sqrt{h_{1}^{-2w}+h_{2}^{-2w}+2h_{1}^{-w}h_{2}^{-w}\cos(\theta_{2}-\theta_{1})} \nonumber .
\end{eqnarray}

Using the definition of the inner product of vectors, we obtain
\begin{eqnarray}
\left\{
\begin{array}{lcl}
\cos(\theta_{1}-\theta) = \hat{\mathbf{V}}\cdot\hat{\mathbf{h}}_{1}=-\frac{h_{1}^{-w}+h_{2}^{-w}\cos(\theta_{2}-\theta_{1})}{\sqrt{h_{1}^{-2w}+h_{2}^{-2w}+2h_{1}^{-w}h_{2}^{-w}\cos(\theta_{2}-\theta_{1})}},\\
\cos(\theta_{2}-\theta) =\hat{\mathbf{V}}\cdot\hat{\mathbf{h}}_{2} = -\frac{h_{2}^{-w}+h_{1}^{-w}\cos(\theta_{2}-\theta_{1})}{\sqrt{h_{1}^{-2w}+h_{2}^{-2w}+2h_{1}^{-w}h_{2}^{-w}\cos(\theta_{2}-\theta_{1})}}. \nonumber
\end{array}
\right.
\end{eqnarray}

As a result, we have
\begin{eqnarray}
\left\{
\begin{array}{lcl}
\cos\theta = -\frac{h_{1}^{-w}\sin\theta_{2}+h_{2}^{-w}\cos(\theta_{2}-\theta_{1})\sin\theta_{2}-h_{2}^{-w}\sin\theta_{1}-h_{1}^{-w}\cos(\theta_{2}-\theta_{1})\sin\theta_{1}}{\sqrt{h_{1}^{-2w}+h_{2}^{-2w}+2h_{1}^{-w}h_{2}^{-w}\cos(\theta_{2}-\theta_{1})}\sin(\theta_{2}-\theta_{1})},\\
\sin\theta = -\frac{h_{1}^{-w}\cos\theta_{2}+h_{2}^{-w}\cos(\theta_{2}-\theta_{1})\cos\theta_{2}-h_{2}^{-w}\cos\theta_{1}-h_{1}^{-w}\cos(\theta_{2}-\theta_{1})\cos\theta_{1}}{\sqrt{h_{1}^{-2w}+h_{2}^{-2w}+2h_{1}^{-w}h_{2}^{-w}\cos(\theta_{2}-\theta_{1})}\sin(\theta_{1}-\theta_{2})}. \nonumber
\end{array}
\right.
\end{eqnarray}

Hence, the dynamical equations of the motion for defector $D$ are given by
\begin{eqnarray}
\left\{
\begin{array}{lcl}
\dot{x}_{D} = -v\frac{h_{1}^{-w}\sin\theta_{2}+h_{2}^{-w}\cos(\theta_{2}-\theta_{1})\sin\theta_{2}-h_{2}^{-w}\sin\theta_{1}-h_{1}^{-w}\cos(\theta_{2}-\theta_{1})\sin\theta_{1}}{\sqrt{h_{1}^{-2w}+h_{2}^{-2w}+2h_{1}^{-w}h_{2}^{-w}\cos(\theta_{2}-\theta_{1})}\sin(\theta_{2}-\theta_{1})},\\
\dot{y}_{D} =-v\frac{h_{1}^{-w}\cos\theta_{2}+h_{2}^{-w}\cos(\theta_{2}-\theta_{1})\cos\theta_{2}-h_{2}^{-w}\cos\theta_{1}-h_{1}^{-w}\cos(\theta_{2}-\theta_{1})\cos\theta_{1}}{\sqrt{h_{1}^{-2w}+h_{2}^{-2w}+2h_{1}^{-w}h_{2}^{-w}\cos(\theta_{2}-\theta_{1})}\sin(\theta_{1}-\theta_{2})}.
\end{array}
\right.
\end{eqnarray}
where $h_{1}=\sqrt{x_{D}^{2}+(y_{D}-y_{1}+vt)^2}$, $h_{2}=\sqrt{x_{D}^{2}+(y_{D}-y_{2}-vt)^2}$, $\cos\theta_{i}=\frac{x_{D}}{\sqrt{x_{D}^{2}+(y_{D}-y_{C_{i}})^2}}(i=1,2)$, and $\sin\theta_{i}=\frac{y_{D}-y_{C_{i}}}{\sqrt{x_{D}^{2}+(y_{D}-y_{C_{i}})^2}}$.

According to the above equations, we can further calculate the average distance $d_{CD}$ between cooperators $C_1$, $C_2$ and defector $D$ for $\beta=0$ in this simplified motion model. To do that, we solve Eq.~(B.2) via numerical integrations by using Runge-Kutta method \cite{Press_cup92} with time step $dt=10^{-3}$. The initial conditions are $h_{2}(0) = 2$, $w=2$, $-\pi/2 < \theta_1(0)<0$, $0 < \theta_2(0)<\pi/2$, and $1\leq h_1(0)\leq 2$. Then we can respectively obtain the $h_1$ and $h_2$ values, and correspondingly the critical distance $d_{CD} = (h_{1}+h_{2})/2$. We emphasize that for each initial value $h_1(0)$, we can obtain a $d_{CD}$ value for fixed $\theta_1(0)$ value, and the average distance $d_{CD}$ for $\beta=0$ in Fig.~\ref{fig5}(b) is obtained by averaging over all these distance values for uniformly distributed initial values $\theta_1(0)$ between $(-\pi/2, 0)$ and uniformly distributed initial values $\theta_2(0)$ between $(0, \pi/2)$.

\section*{B2. The case of $\beta=1$}
\begin{figure}
\setcounter{figure}{0}
\renewcommand\thefigure{{B2}}
\centering
\includegraphics[width=3.0 in]{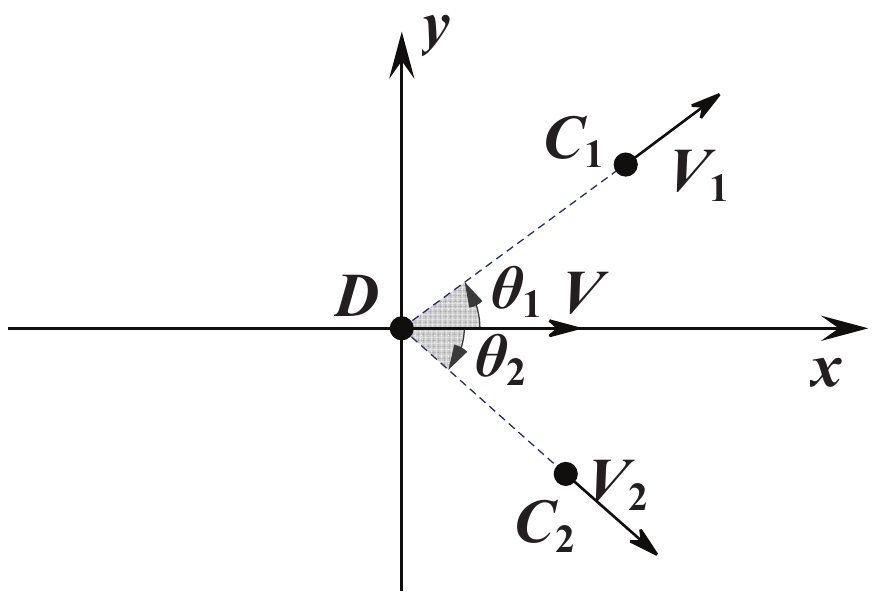}
\caption{ Two cooperators escape away from one defector. $\theta_{1}$ ($\theta_{2}$) represents the angle between the vector velocities of defector $D$ and cooperator $C_{1}$ ($C_2$).}
\label{FigB2}
\end{figure}

In this case, we know that two cooperators escape away from defector $D$. Since there are no other defector in the neighborhood of defector $D$, we consider that defector $D$ moves along with a fixed direction, which is the direction of $\mathbf{v}$. For simplicity, we consider that the migration direction of defector $D$ is the positive direction of $x$-axis in the cartesian coordinate (see Fig.~\ref{FigB2}), and correspondingly set the dynamical position of defector $D$ as $(vt, 0)$. Hence we have $h_{i}=\sqrt{(vt-x_{C_{i}})^{2}+y^2_{C_{i}}}$.
We consider that
$x_{C_1}(0)=h_1(0)\cos\theta_1(0)$ and $y_{C_1}(0)=h_1(0)\sin\theta_1(0)$ for cooperator $C_1$, and $x_{C_2}(0)=h_2(0)\cos\theta_2$ and $y_{C_2}(0)=h_2(0)\sin\theta_2(0)$ for cooperator $C_2$.

Meanwhile, for cooperators $C_{1}$ and $C_{2}$ the dynamical equations of motion can be described as
\begin{eqnarray}
\left\{
\begin{array}{lcl}
\dot{x}_{C_{i}} = v\cos\theta_{i},\\
\dot{y}_{C_{i}} = v\sin\theta_{i},
\end{array}
\right.
\end{eqnarray}
where $\theta_{i}(i=1,2)$ represent the angle between the vector velocities of defector $D$ and cooperator $C_{i}$.

In addition, we have $\tan\theta_{i}=\frac{y_{C_{i}}}{x_{C_{i}}-vt}$. Accordingly, we have
\begin{eqnarray}
\tan^{2}\theta_{i}=\frac{y^2_{C_{i}}}{(x_{C_{i}}-vt)^{2}},  \nonumber
\end{eqnarray}
and
\begin{eqnarray}
\frac{1}{\cos^{2}\theta_{i}}=1+\frac{y^2_{C_{i}}}{(x_{C_{i}}-vt)^{2}}. \nonumber
\end{eqnarray}

Considering that $\theta_{i}$ should be restricted between $(0, \pi/2)$ or $(-\pi/2, 0)$, thus we have
\begin{eqnarray}
\cos\theta_{i}=\frac{x_{C_{i}}-vt}{\sqrt{(x_{C_{i}}-vt)^{2}+y^2_{C_{i}}}}, \nonumber
\end{eqnarray}
and
\begin{eqnarray}
\sin\theta_{i}=\frac{y_{C_{i}}}{\sqrt{(x_{C_{i}}-vt)^{2}+y^2_{C_{i}}}}.  \nonumber
\end{eqnarray}

Hence, the dynamical equations of motion for two cooperators become
\begin{eqnarray}
\left\{
\begin{array}{lcl}
\dot{x}_{C_{i}} = \frac{vx_{C_{i}}-v^2t}{\sqrt{(x_{C_{i}}-vt)^{2}+y^2_{C_{i}}}},\\
\dot{y}_{C_{i}} = \frac{vy_{C_{i}}}{\sqrt{(x_{C_{i}}-vt)^{2}+y^2_{C_{i}}}}.
\end{array}
\right.
\end{eqnarray}

According to these equations, we can further calculate the average distance $d_{CD}$ between cooperators $C_1$, $C_2$ and defector $D$ for $\beta=1$ in this simplified motion model. To do that, we solve Eq.~(B.4) via numerical integrations by using Runge-Kutta method \cite{Press_cup92} with time step $dt=10^{-3}$. The initial conditions are $h_{2}(0) = 2$, $0 < \theta_1(0)<\pi/2$, $-\pi/2< \theta_2(0)<0$, and $1\leq h_1(0)\leq 2$. Then we can respectively obtain the $h_1$ and $h_2$ values, and correspondingly have $d_{CD} = (h_{1}+h_{2})/2$. We emphasize that for each initial value $h_1(0)$, we can obtain a $d_{CD}$ value for fixed $\theta_1(0)$ and $\theta_2(0)$ values, and the average distance $d_{CD}$ for $\beta=1$ in Fig.~\ref{fig5}(b) is obtained by averaging over all these distance values for uniformly distributed initial values $\theta_1(0)$ between $(0, \pi/2)$ and uniformly distributed initial values $\theta_2(0)$ between $(-\pi/2, 0)$.

\end{document}